\documentclass[onecolumn,12pt]{IEEEtran}
\usepackage{algorithmic}
\usepackage{amsmath}
\usepackage{latexsym}
\usepackage{epsfig}
\usepackage{amsmath}
\usepackage{multicol}
\usepackage{amssymb}
\usepackage{subfigure}
\usepackage{cite}
\usepackage{setspace}

\def\C{{\rm C}}

\newcommand{\qed}{\nobreak \ifvmode \relax \else
      \ifdim\lastskip<1.5em \hskip-\lastskip
      \hskip1.5em plus0em minus0.5em \fi \nobreak
      \vrule height0.75em width0.5em depth0.25em\fi}
\title{Coordinated Direct and Relay Schemes for Two-Hop Communication in VANETS}
\author{\begin{tabular}{c}
Chan Dai Truyen Thai and Marion Berbineau\\
Univ Lille Nord de France, IFSTTAR, LEOST, Villeneuve d'Ascq\\
\end{tabular}}

\begin{document}
\maketitle

\begin{abstract}
In order to accommodate increasing need and offer communication with high performance, both vehicle-to-infrastructure (V2I) and vehicle-to-vehicle (V2V) communications are exploited. The advantages of static nodes and vehicular nodes are combined to achieve an optimal routing scheme. In this paper, we consider the communications between a static node and the vehicular nodes moving in an adjacent area of it. The adjacent area is defined as the zone where a vehicular can communicate with the static node within maximum two hops. We only consider single-hop and two-hop transmissions because these transmissions can be considered as building blocks to construct transmissions with a higher number of hops. Different cases in which an uplink or a downlink for the two-hop user combined with an uplink or a downlink for the single-hop user correspond to different CDR schemes. Using side information to intentionally cancel the interference, Network Coding (NC), CDR, overhearing and multi-way schemes aggregate communications flows in order to increase the performance of the network. We apply the mentioned schemes to a V2I network and propose novel schemes to optimally arrange and combine the transmissions.
\end{abstract}

\section{Introduction}
Vehicular ad hoc networks (VANETs) have attracted a great attention in research community due to their huge benefits in safety and entertaining applications \cite{vanet_application}. In order to accommodate increasing need and offer communication with high performance, both vehicle-to-infrastructure (V2I) and vehicle-to-vehicle (V2V) communications are exploited \cite{v2v_v2i}. The advantages of static nodes and vehicular nodes are combined to achieve an optimal routing scheme \cite{routing_survey}.

Considering a two-tier network as such, a message from a vehicular source to its destination may be routed via a certain static node. This static node may store the message until it has a good link to a vehicular node on the way to the destination \cite{sadv}. This static node can also be connected to the backbone network through a backhaul link using an orthogonal channel from over which the message will be forwarded. In another example, a certain message destined to a vehicular node which is nearby that static node can be routed from the source via this static node. To a certain extent, a static node together with its surrounding vehicular nodes can be considered as a small cell. In this paper, we consider the communications between a static node and the vehicular nodes moving in an adjacent area of it. 

\begin{figure}
\centering
\includegraphics[width=0.6\columnwidth]{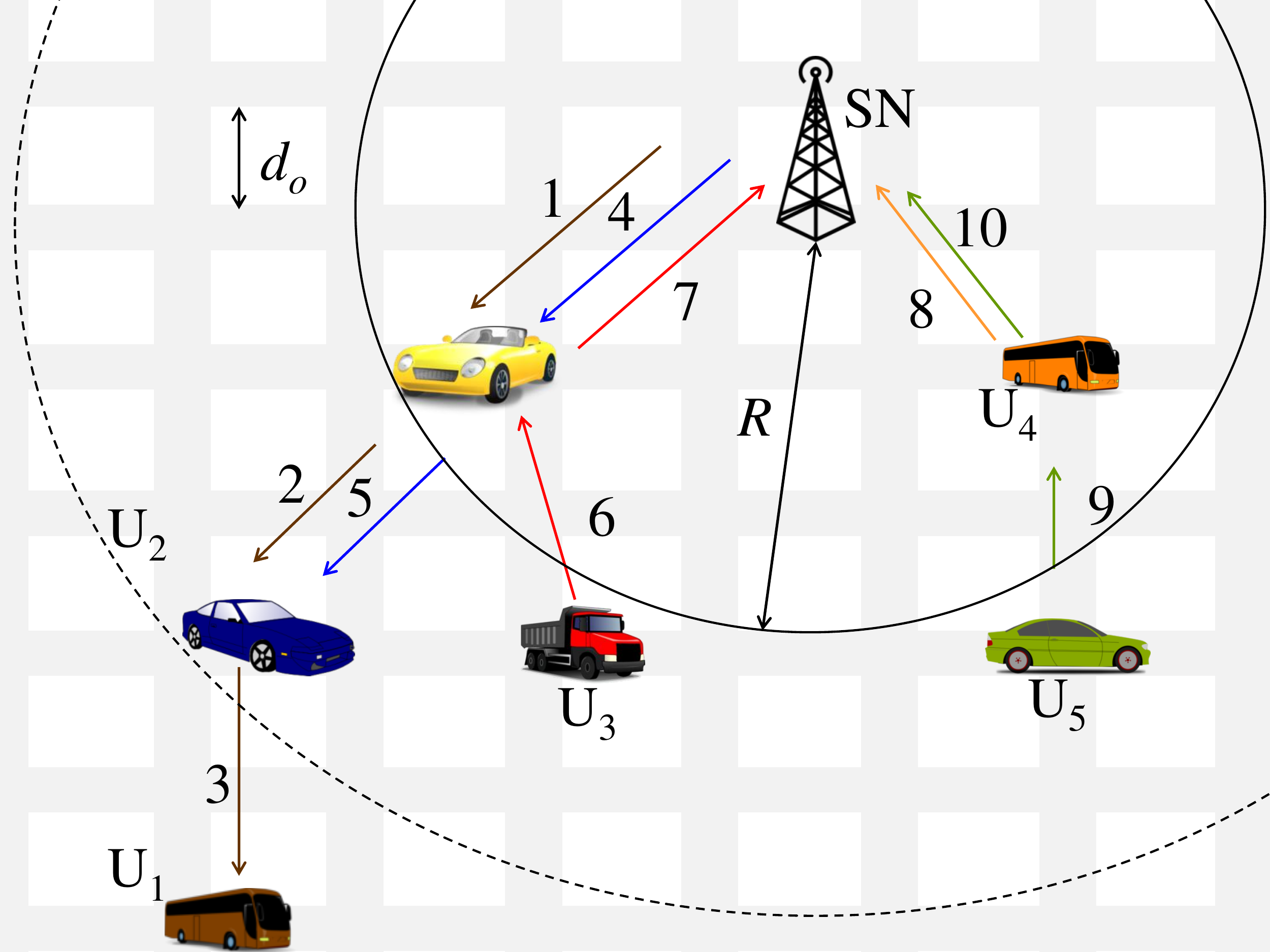}
\caption{We consider vehicular users which can communicate with the SN in at most two hops.}
\label{network}
\end{figure}

The adjacent area is defined as the zone, as shown in Fig. \ref{network}, where a vehicular can communicate with the static node within maximum two hops. We only consider single-hop and two-hop transmissions because these transmissions can be considered as building blocks to construct transmissions with a higher number of hops. Assume at a certain time, we have only a downlink for node U$_1$ and an uplink for node U$_4$. This means that transmissions 1, 2, 3 and 8 are to be conducted. A conventional scheme will conduct all of them in different time slots therefore they do not interfere with each other. In an advanced scheme, two or more transmissions may be conducted simultaneously in order to use less time slots therefore increase the spectrum efficiency. If we only consider transmissions 1, 2 and 8 and try to combine these flows, we have a Coordinated Direct and Relay (CDR) scheme. This CDR scheme combined with transmission 3 can accommodate the required downlink and uplink for U$_1$ and U$_4$ respectively. Different cases in which an uplink or a downlink for the two-hop user combined with an uplink or a downlink for the single-hop user correspond to different CDR schemes which are described in details in \cite{multiflow_chanthai}.

If at a certain time, there are an uplink and a downlink for two users using the same relay node (e.g. U$_2$ and U$_3$ in Fig. \ref{network}) respectively. The transmissions can be combined in an overhearing scheme \cite{overhearing_fansun_comlett,dof_fansun_spawc}. On the other hand, if there are two-hop users using different relay nodes: one has an uplink and one has a downlink (e.g. U$_2$ and U$_5$), we have a multi-way scheme \cite{fourway_huaping_spawc, fourway_huaping_arxiv, fourway_huaping_lett}. If we use the CDR, overhearing and multi-way schemes as described above to respective cases instead of the corresponding conventional schemes, several time slots are saved and a significant improvement is gained.

Using side information to intentionally cancel the interference, Network Coding (NC), CDR, overhearing and multi-way schemes aggregate communications flows in order to increase the performance of the network. Let us regard these types of schemes as \textit{multi-flow schemes}. So far, multi-flow schemes have been proposed and analyzed in terms of spectrum efficiency \cite{multiflow_chanthai} and diversity \cite{dmt_chanthai}. In those works, the nodes are assumed to be static in the whole scheme which may last for several time slots. For a vehicular network, the positions of the nodes in different time slots are different. This may make the performance of the network lower than that of a non-multi-flow based network. In this paper, we apply the mentioned schemes to a V2I network and propose novel schemes to optimally arrange and combine the transmissions.
\section{System Model}\label{system_model}
We consider a static nodes (SN) and $n$ adjacent vehicular nodes U$_i$, $i\in\{1, 2, ..., n\}$. We assume that all considered traffic from/to the vehicular nodes goes through the SN as an intermediate node on the way to the destination. The SN therefore can be considered as a Base Station of a cell in cellular networks. The vehicular nodes with distances to the SN smaller than $R$ can directly communicate with the SN while the vehicular nodes outside the circle with radius $R$ cannot directly communicate with the SN, due to the negligible magnitude of the channel between it and the SN, and have to rely on vehicular nodes inside the circle as relays.

Consider a map with horizontal and vertical streets equally separated with distance $d_o$. At time $t$, vehicular node $i$ moves with velocity $v_i$ to the direction $d_i$, where $d_i = 1, 2, 3, 4$ correspond to East, North, West, South. At an intersection, it turns left, goes straight or turns right with probabilities $\frac{1 - p_o}2$, $p_o$ and $\frac{1 - p_o}2$ respectively. Assume that at the beginning of a scheme which lasts for a few time slots, all vehicular nodes in the map reports their expected positions in the whole scheme to the SN. The positions of a vehicle in a few time slots can be calculated.

All transmissions are in one frequency with a normalized bandwidth of 1 Hz. All nodes are single-antenna and half-duplexed. Each of the complex channels is reciprocal, known at the receivers. The receivers here include overhearing receivers as well as receivers of the second-hop transmission of an Amplify-and-Forward (AF) relaying communication. Each vehicular node requests an uplink or a downlink transmission to the SN. We assume that the data to/from each user is \emph{infinitely backlogged} so that there are always data to transmit as in many works regarding downlink \cite{backlog3} and Two-way Relaying optimization \cite{backlog2} and scheduling \cite{pf1,backlog}. Thus the achievable rate for a user at a certain time is equal to the information theoretic capacity, i.e. $\C(\gamma) = \log_2(1 + \gamma)$, where $\gamma$ denotes an instantaneous received SINR at the receiver.

We use the following notation, with a slight abuse: $x_i$ may denote a packet or a single symbol, and it will be clear from the context. For example, the packet that the SN wants to send to a user is denoted by $x$; but if we want to express the signal received, then we use expressions of type $y = hx + z$, where $y$, $x$ and $z$ denote symbols (received, sent and noise respectively). We assume perfect power control i.e. the transmit power is selected so as the received power at the aimed receiver is at a fixed level of $P$ \cite{femto_uplink}. The principle is also applied to the case of AF or DF transmission. The received power at the relay and at the final receiver is also $P$.

The considered scheme is divided into $2n$ time slots so that each two of them can be assigned to each user. If it is a relayed user, each time slot is used for each of two hop transmissions. If it is a direct user, only one of the two slots assigned to it is used. The number of total slots used is $n_T = 2n_1 + n_2$, where $n_1$ and $n_2$ are the numbers of relayed and direct users respectively. Note that $n_T$ depends on the way the transmissions of all users are scheduled because a user which is relayed in a certain slot can be direct in another time slot because it moves toward the SN.
\section{Scheduling}\label{Scheduling}
\begin{figure}
\centering
\includegraphics[width=0.5\columnwidth]{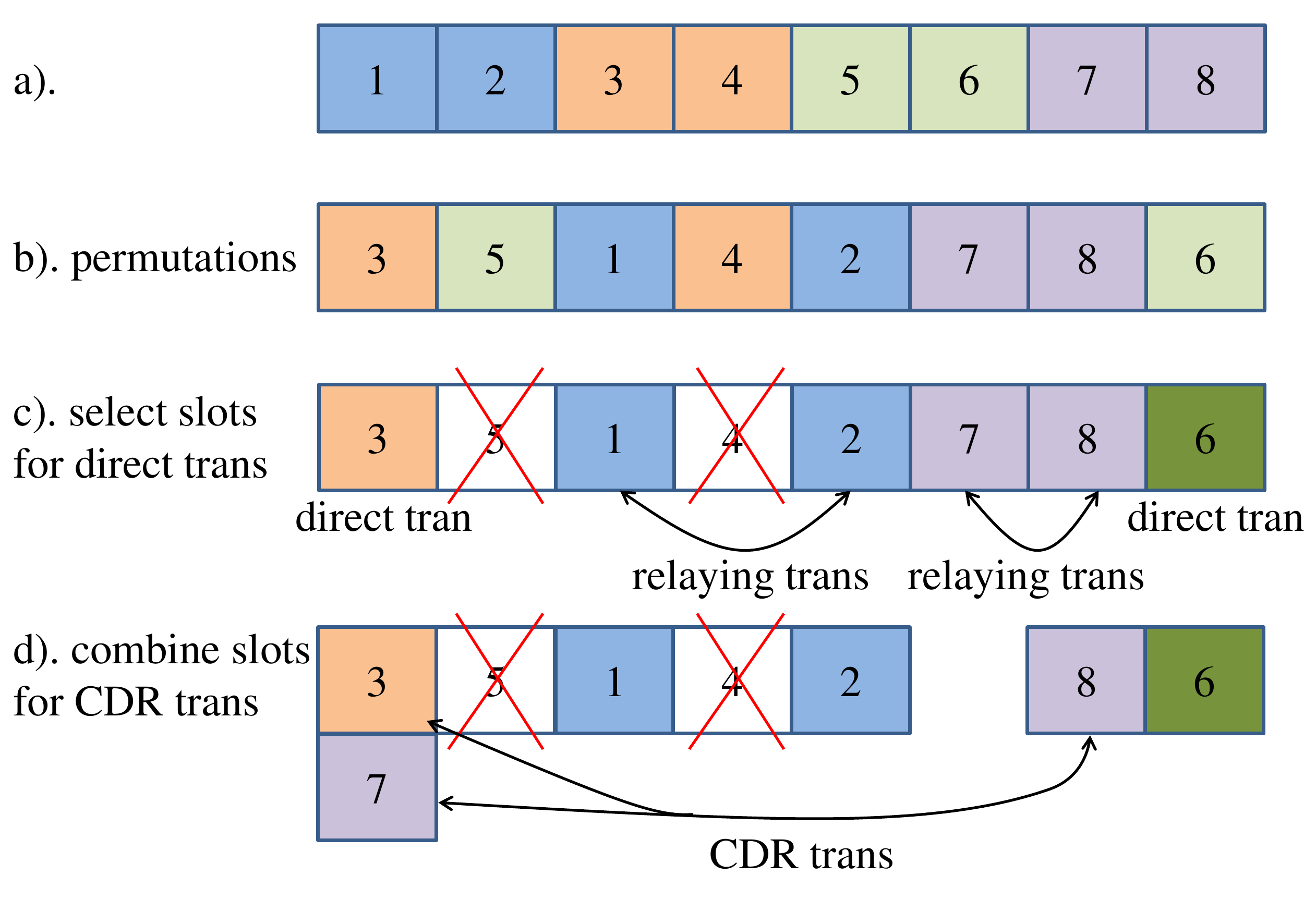}
\caption{Optimally scheduling by permuting, selecting time slots and combining transmissions for the case with $n = 4$ users.}
\label{scheduler}
\end{figure}
\subsection{Single-Flow Scheme}
The single-flow scheme is a conventional scheme in which, a direct communication requires one time slot while a relay communication requires two slots. The purpose is to schedule the transmissions so that the performance is maximized considering a fixed consumed energy. For a direct user, it is optimal when it is near the SN. For a relayed user, it should be when the relay moves in a good direction e.g. the relay moves from the relayed user toward the SN when there is an uplink. Optimality of a user does not mean optimality of the whole network therefore we will find the optimal transmission scheduling by trying all the permutations of all transmissions to see which scheduling gives the highest performance.

Assigning time slots $2i - 1$ and $2i$ to user $i$, we have totally $2n$ slots as in Fig. \ref{scheduler}a. If user $i$ is a relayed user, slots $2i - 1$ and $2i$ are used for first-hop and second-hop transmissions respectively. First, from the list of the slots in the first row, all permutations of the slots are listed. There are totally $(2n)!$ permutations. The permutations in which slot $2i$ appears before slot $2i - 1$ is invalid and therefore crossed out. Fig. \ref{scheduler}b shows one of the valid permutation.

In a permutation, the two slots assigned to each user is checked. If the distance from a user to the SN is smaller than $R$ ($d(t) \leq R$) in only one of the two slots, that slot is used for the user. For example, Fig. \ref{scheduler}c shows that between slots 3 and 4 of user 2, slot 4 is worse and slot 3 is chosen. For user 3, slot 4 is chosen rather than slot 6. If in both slots $d(t) \leq R$, the slot when it is closer to the SN is selected. If in both slots $d(t) > R$, the two slots are used for the first-hop and second-hop transmissions respectively as slots 1 and 2 of user 1, slots 7 and 8 of user 4. The rate for each user is calculated. The sum--rate for all users in one permutation is calculated. The permutation with the highest sum--rate is selected.
\subsection{Multi-Flow Scheme} 
After the transmissions are fixed in the single-flow scheme, we still can combine transmissions to decrease the time slots used in order to increase the spectrum efficiency. The combination is performed based on the optimal permutation of the single-flow scheme therefore the energy consumptions of the single-flow and multi-flow schemes are almost the same. The advanced schemes we applied here include CDR, overhearing and multi-way schemes.
\subsubsection{CDR schemes} 
First, we look for a direct user and a relayed user which can be combined in two certain time slots. In one slot, the direct transmission and a relayed transmission (can be the first or the second hop depending on which CDR scheme is used) are conducted simultaneously. Let us call this one the CDR simultaneous slot as slots 3 and 7 in Fig. \ref{scheduler}d. In the other slot called CDR single slot, the other relayed transmissions is conducted as slot 8 in the figure.

Second, we look for a time slot which can host slots 3 and 7 by step by step move slots 3 and 7 along the row and check if it fits (the slot must be empty and the direct user is still direct in the slot) and calculate the performance metric in that case. The case with the highest performance is selected.
\section{Description of Individual Schemes}\label{schemes_description}
\begin{figure*}
\centering
\includegraphics[width=1\columnwidth]{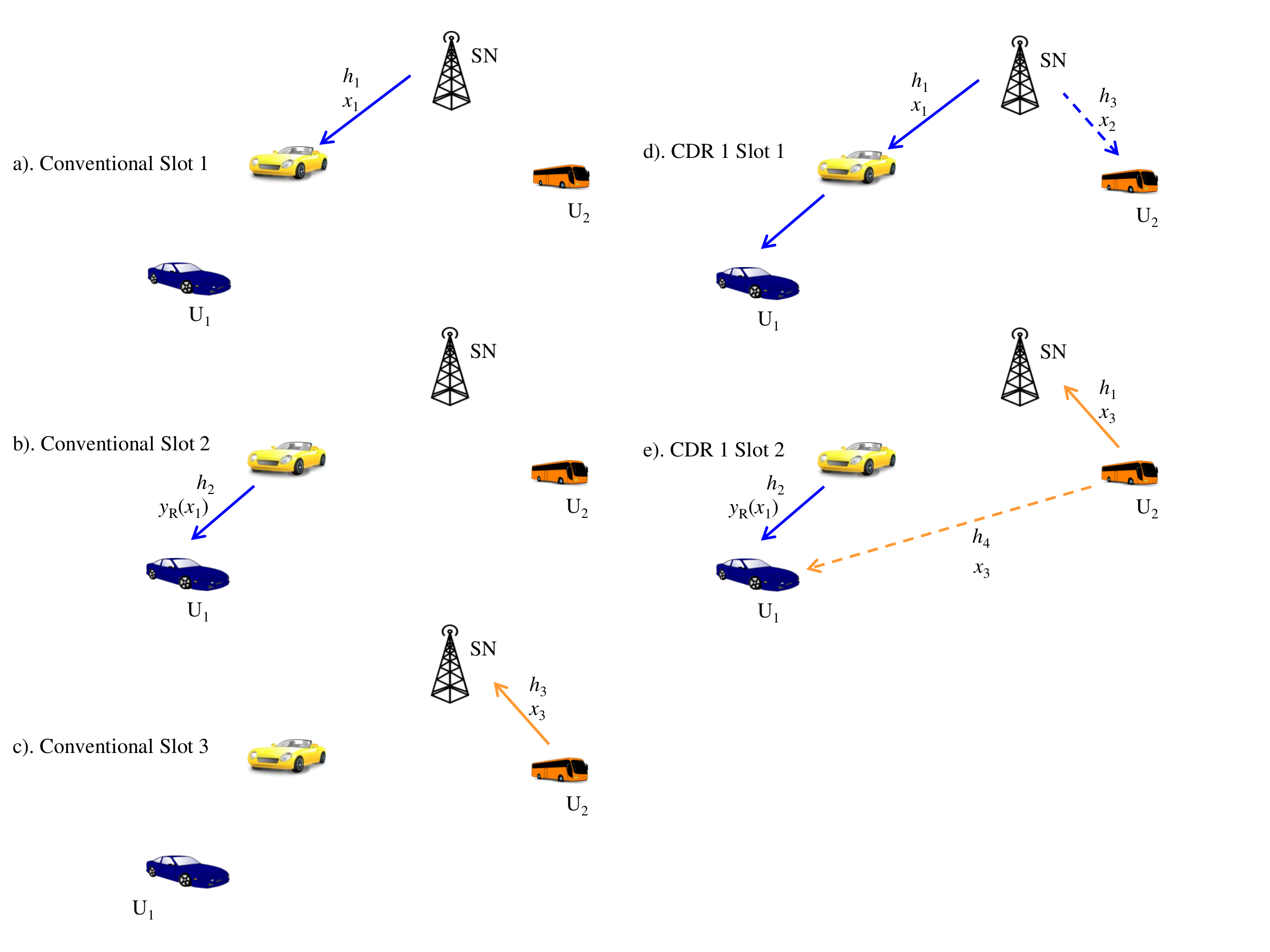}
\caption{Single-flow (conventional) and multi-flow (CDR 1) individual schemes.}
\label{cdr1}
\end{figure*}

In this section, we present single-flow (conventional) and multi-flow (CDR) individual schemes in two relaying modes: AF and DF. An individual scheme is presented for two users: one relayed user (denoted as user 1) and one direct user (user 2). The vehicular node acts as a relay is denoted as R as shown in Fig. \ref{cdr1}. The whole composite scheme is a multiplexing of $\frac n2$ individual schemes. 

The signals for user 1 and 2 are $x_1$ and $x_2$ respectively. We denote the received signal, noise and channel in time slot 1, 2 and 3 as $y,z,h$, $y',z',h'$, $y'',z'',h''$ respectively. Denote $_i\mathrm{SNR}^j_k$ as the SNR when signal $x_k$ is decoded at node $i$ in decoding option $j$.
\subsection{Amplify-and-Forward}
\subsubsection{Non-CDR Schemes}
The scheme is conducted in three equal time slots. In time slot 1, the SN transmits $x_1$ towards the relay. The transmit power is set to $\frac P{|h_1|^2}$ so as to compensate the SN-to-relay channel. Because channel between the SN and user 1 has a negligible magnitude, user 1 does not receive the signal from the SN. The relay receives
\begin{equation}
y_R = \frac{h_1\sqrt{P}}{|h_1|}x_1 + z_R,
\end{equation}
and scales with a power factor
\begin{equation}
\alpha = \frac 1{P + \sigma^2}.
\end{equation}

In time slot 2, the relay transmits the amplified version of the received signal in time slot 1 towards user 1. User 1 receives
\begin{equation}\begin{array}{lll}
y_1' &=& \frac{h_2'\sqrt{\alpha P}}{|h_2'|}y_R + z_1'\\
&=&
\frac{h_2'\sqrt{\alpha P}}{|h_2'|}\frac{h_1\sqrt{P}}{|h_1|}x_1 + \frac{h_2'\sqrt{\alpha P}}{|h_2'|}z_R + z_1'
\end{array}\end{equation}
and decode with SNR
\begin{equation}
_1\mathrm{SNR}_1= \frac{\alpha P^2}{\alpha P\sigma^2 + \sigma^2} = \frac{\gamma_o^2}{2\gamma_o + 1}.
\end{equation}

In time slot 3, user 2 transmits $x_2$ to the SN. The SN receives
\begin{equation}
y''_S = \frac{h''_3\sqrt{P}}{|h''_3|}x_1 + z''_S,
\end{equation}
and decodes with SNR
\begin{equation}
_S\mathrm{SNR}_2= \gamma_o.
\end{equation}

Finally, we have the rates for two users
\begin{equation}
\left\{\begin{array}{l}R_1 \leq \C\left(_1\mathrm{SNR}_1\right)\\R_2 \leq \C\left(_S\mathrm{SNR}_2\right)\end{array}\right.
\end{equation}
\subsubsection{CDR Schemes} 
The scheme is conducted in two equal time slots. In time slot 1, the SN transmits $x_1$ towards the relay. The transmit power is set to $\frac P{|h_1|^2}$ so as to compensate the SN-to-relay channel. Because channel between the SN and user 1 has a negligible magnitude, user 1 does not receive the signal from the SN. The relay receives
\begin{equation}
y_R = \frac{h_1\sqrt{P}}{|h_1|}x_1 + z_R,
\end{equation}
and scales with a power factor
\begin{equation}
\alpha = \frac 1{P + \sigma^2}.
\end{equation}

In time slot 2, the relay transmits the amplified version of the received signal in time slot 1 towards user 1 and user 2 transmits $x_2$ towards the SN simultaneously. The transmit power of the first transmission is set to $\frac{\alpha P}{|h_2|^2}$ so that user 1 receives the signal with power of $P$. User 1 and the SN respectively receive
\begin{equation}\begin{array}{lll}
y_1' &=& \frac{h_2'\sqrt{\alpha P}}{|h_2'|}y_R + \frac{h_4'\sqrt{P}}{|h_3'|}x_2 + z_1'\\
&=&
\frac{h_2'\sqrt{\alpha P}}{|h_2'|}\frac{h_1\sqrt{P}}{|h_1|}x_1 + \frac{h_2'\sqrt{\alpha P}}{|h_2'|}z_R + \frac{h_4'\sqrt{P}}{|h_3'|}x_2 + z_1',
\end{array}\end{equation}
\begin{equation}\begin{array}{lll}
y_F' &=& \frac{h_1'\sqrt{\alpha P}}{|h_2'|}y_R + \frac{h_3'\sqrt{P}}{|h_3'|}x_2 + z_F'\\
&=&
\frac{h_1'\sqrt{\alpha P}}{|h_2'|}\frac{h_1\sqrt{P}}{|h_1|}x_1 + \frac{h_1'\sqrt{\alpha P}}{|h_2'|}z_R + \frac{h_3'\sqrt{P}}{|h_3'|}x_2 + z_F'.
\end{array}\end{equation}
Because the SN has the information about $x_1$ and the channels, the contribution of $x_1$ in $y_F'$ is cancelled
\begin{equation}\begin{array}{lll}
\tilde{y}_S' = \frac{h_1'\sqrt{\alpha P}}{|h_2'|}z_R + \frac{h_3'\sqrt{P}}{|h_3'|}x_2 + z_F'.
\end{array}\end{equation}

At user 1, there are two options to decode
\begin{itemize}
\item Option 1: User 1 decodes $x_1$ treating $x_2$ as noise with SNR, using MMSE \cite{mmse_fansun_siglett1, mmse_fansun_siglett2},
\begin{equation}
_1\mathrm{SNR}^1_1= \frac{\alpha P^2}{\frac{|h_4'|^2}{|h_3'|^2}P + \alpha P\sigma^2 + \sigma^2} = \frac{\gamma_o^2}{\left(\frac{\gamma_4'}{\gamma_3'}\gamma_o + 1\right)(\gamma_o + 1) + \gamma_o}.
\end{equation}
\item Option 2: User 1 decodes $x_2$ treating $x_1$ as noise with SNR
\begin{equation}
_1\mathrm{SNR}^2_2 = \frac{\frac{|h_4'|^2}{|h_3'|^2}P}{\alpha P^2 + \alpha P\sigma^2 + \sigma^2} = \frac{\gamma_4'\gamma_o}{\gamma_3'(\gamma_o + 1)},
\end{equation}
cancels the contribution of $x_2$ in $\tilde{y}_1$ and decodes $x_1$ with SNR
\begin{equation}
_1\mathrm{SNR}^2_1 = \frac{\alpha P^2}{\alpha P\sigma^2 + \sigma^2} = \frac{\gamma_o^2}{2\gamma_o + 1}.
\end{equation}
This option corresponds to the case when the contribution of $x_2$ in $\tilde{y}_1$ is higher than that of $x_1$. In the opposite case, option 1 is appropriate.
\end{itemize}

In both options, the SN decodes $x_2$ from $\tilde{y}_S$ with
\begin{equation}
_S\mathrm{SNR}_2 = \frac{P}{\frac{|h_1'|^2\alpha P}{|h_2'|^2}\sigma^2 + \sigma^2} = \frac{\gamma_o}{\frac{\gamma_1'}{\gamma_2'} + \gamma_o + 1}.
\end{equation}
In summary, we have two options
\begin{equation}
\left\{\begin{array}{l}R_1 \leq \C\left(_1\mathrm{SNR}_1^1\right)\\R_2 \leq \C\left(_S\mathrm{SNR}_2\right)\end{array}\right.
\mbox{or}
\left\{\begin{array}{l}R_1 \leq \C\left(_1\mathrm{SNR}_1^2\right)\\R_2 \leq \C\left(\min\left(_1\mathrm{SNR}_2^2, _S\mathrm{SNR}_2\right)\right)\end{array}\right.
\end{equation}
\subsection{Decode-and-Forward} 
\subsubsection{Non-CDR Schemes}
The scheme is conducted in three equal time slots. In time slot 1, the SN transmits $x_1$ towards the relay. The transmit power is set to $\frac P{|h_1|^2}$ so as to compensate the SN-to-relay channel. Because channel between the SN and user 1 has a negligible magnitude, user 1 does not receive the signal from the SN. The relay receives
\begin{equation}
y_R = \frac{h_1\sqrt{P}}{|h_1|}x_1 + z_R,
\end{equation}
and scales decodes $x_1$ with SNR $_R\mathrm{SNR}_1 = \frac{P}{\sigma^2} = \gamma_o$, transmits $x_1$ in time slot 2 towards user 1. User 1 receives and decodes with SNR $_1\mathrm{SNR}_1 = \gamma_o$. In time slot 3, user 2 transmits $x_2$ to the SN. The SN receives and decodes with SNR $_S\mathrm{SNR}_1 = \gamma_o$.

Finally, we have the rates for two users
\begin{equation}
\left\{\begin{array}{l}R_1 \leq \C(\gamma_o)\\R_2 \leq \C(\gamma_o).\end{array}\right.
\end{equation}
\subsubsection{CDR Schemes}
In time slot 1, the relay receives
\begin{equation}
y_R = \frac{h_1\sqrt{P}}{|h_1|}x_1 + z_R,
\end{equation}
and decodes $x_1$ with SNR
\begin{equation}
_1\mathrm{SNR}_1= \gamma_o.
\end{equation}

In time slot 2, the relay transmits $x_1$ and user 2 transmits $x_2$ simultaneously. User 1 and the SN respectively receive
\begin{equation}
y_1' = \frac{h_2'\sqrt{P}}{|h_2'|}x_1 + \frac{h_4'\sqrt{P}}{|h_3'|}x_2 + z_1',
\end{equation}
\begin{equation}
y_F' = \frac{h_1'\sqrt{P}}{|h_2'|}x_1 + \frac{h_3'\sqrt{P}}{|h_3'|}x_2 + z_1',
\end{equation}

\begin{itemize}
\item Option 1: User 1 decodes $x_1$ treating $x_2$ as noise with SNR
\begin{equation}
_1\mathrm{SNR}^1_1= \frac{\gamma_o}{\frac{\gamma_4'}{\gamma_3'}\gamma_o + 1}.
\end{equation}
\item Option 2: User 1 decodes $x_2$ treating $x_1$ as noise with SNR
\begin{equation}
_1\mathrm{SNR}^2_2 = \frac{\gamma_4'\gamma_o}{\gamma_3'(\gamma_o + 1)},
\end{equation}
cancels the contribution of $x_2$ in $\tilde{y}_1$ and decodes $x_1$ with SNR
\begin{equation}
_1\mathrm{SNR}^2_1 = \gamma_o.
\end{equation}
\end{itemize}

In summary, we have two options
\begin{equation}
\left\{\begin{array}{l}R_1 \leq \C\left(\min\left(_1\mathrm{SNR}_1^1, _R\mathrm{SNR}_1\right)\right)\\R_2 \leq \C\left(_S\mathrm{SNR}_2\right)\end{array}\right.
\mbox{or}
\left\{\begin{array}{l}R_1 \leq \C\left(\min\left(_1\mathrm{SNR}_1^2, _R\mathrm{SNR}_1\right)\right)\\R_2 \leq \C\left(\min\left(_1\mathrm{SNR}_2^2, _S\mathrm{SNR}_2\right)\right)\end{array}\right.
\end{equation}

\begin{figure}
\centering
\includegraphics[width=1\columnwidth]{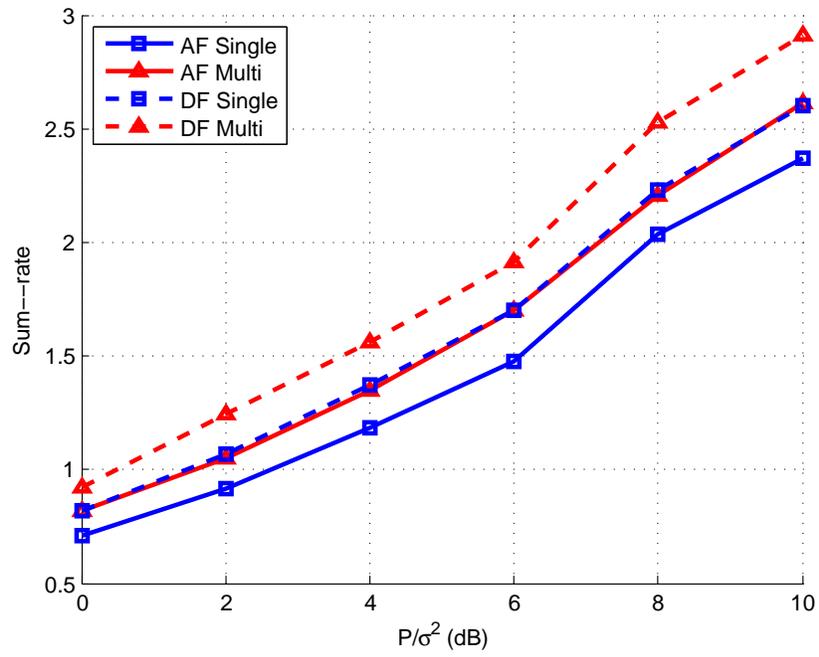}
\caption{Sum--rate versus $\frac P{\sigma^2}$.}
\label{sum_rate_vs_gamma}
\end{figure}
\begin{figure}
\centering
\includegraphics[width=1\columnwidth]{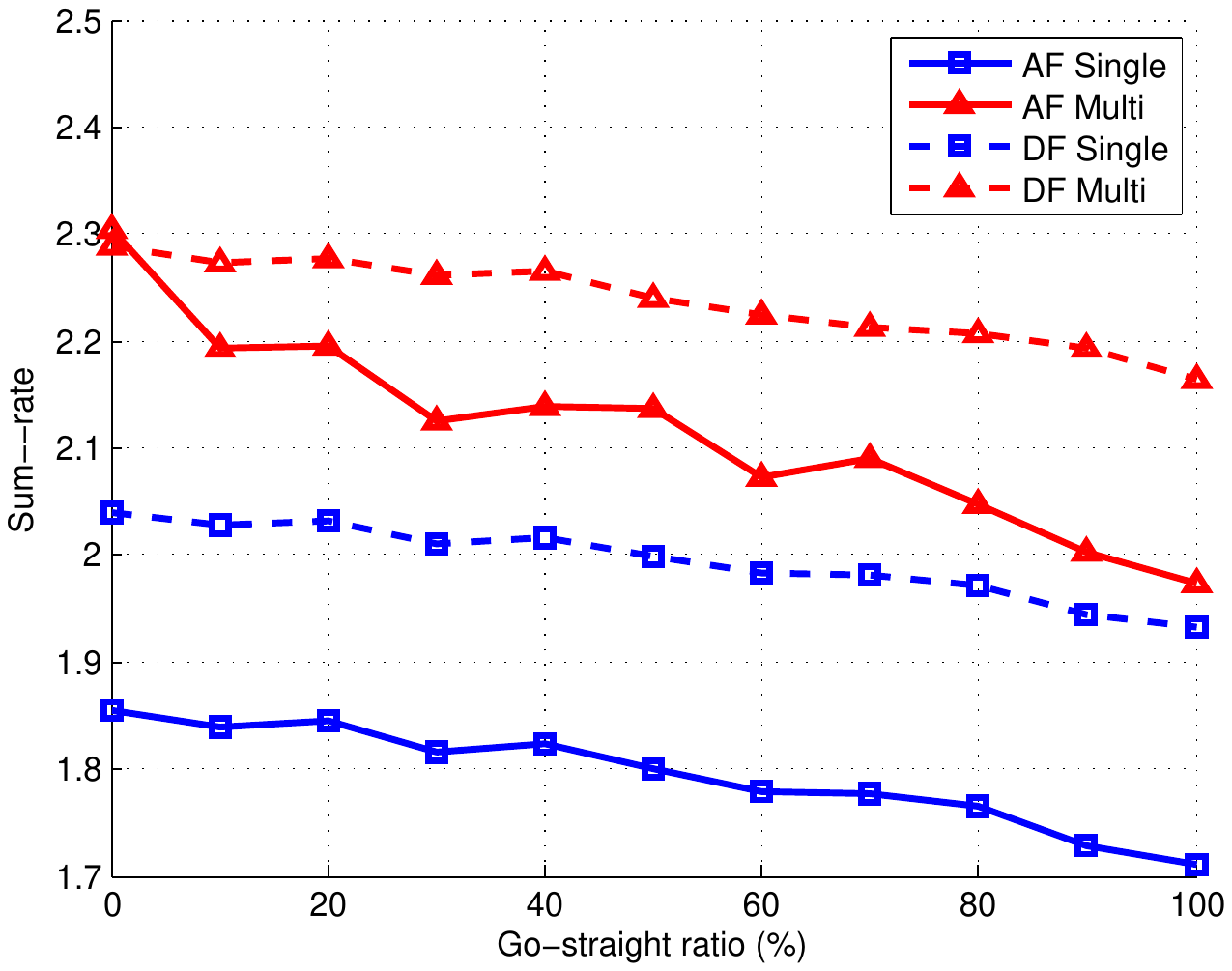}
\caption{Sum--rate versus go-straight ratio $p_o$ (\%).}
\label{sum_rate_vs_go_straight}
\end{figure}
Monto Carlo simulation is conducted for a map with parallel horizontal and vertical streets equally separated with $d_o = 30$m. Other parameters include maximum one-hop distance $R = 65$m, $n = 4$ users. All vehicular nodes move with the same velocity $v = 10m/$time slot.

\appendix[Proof of Proposition]
\subsection{$S_2$}
\begin{equation}
y_2 = \frac{h_4\sqrt{P}}{|h_2|}x_1 + \frac{h_3\sqrt{P}}{|h_3|}x_2 + z_2.
\end{equation}
\begin{equation}
y_R = \frac{h_2\sqrt{P}}{|h_2|}x_1 + \frac{h_1\sqrt{P}}{|h_3|}x_2 + z_R.
\end{equation}
\begin{equation}
\alpha = \frac 1{P + \frac{|h_1|^2}{|h_3|^2}P + \sigma^2}.
\end{equation}
\begin{equation}\begin{array}{lll}
y_F' &=& \frac{h_1'\sqrt{\alpha P}}{|h_1'|}y_R + z_F'\\
&=&
\frac{h_1'\sqrt{\alpha P}}{|h_1'|}\frac{h_2\sqrt{P}}{|h_2|}x_1 + \frac{h_1'\sqrt{\alpha P}}{|h_1'|}\frac{h_1\sqrt{P}}{|h_3|}x_2 + \frac{h_1'\sqrt{\alpha P}}{|h_1'|}z_R + z_F'.
\end{array}\end{equation}
\begin{equation}
\tilde{y}_S' = \frac{h_1'\sqrt{\alpha P}}{|h_1'|}\frac{h_2\sqrt{P}}{|h_2|}x_1 + \frac{h_1'\sqrt{\alpha P}}{|h_1'|}z_R + z_F'.
\end{equation}
\begin{equation}
_S\mathrm{SNR}_1 = \frac{\alpha P^2}{\alpha P\sigma^2 + \sigma^2}.
\end{equation}
\begin{equation}\begin{array}{lll}
y_2' &=& \frac{h_5'\sqrt{\alpha P}}{|h_1'|}y_R + z_1'\\
&=&
\frac{h_5'\sqrt{\alpha P}}{|h_1'|}\frac{h_2\sqrt{P}}{|h_2|}x_1 + \frac{h_5'\sqrt{\alpha P}}{|h_1'|}\frac{h_1\sqrt{P}}{|h_3|}x_2 + \frac{h_5'\sqrt{\alpha P}}{|h_1'|}z_R + z_2',
\end{array}\end{equation}
\begin{equation}
\mathbf{H} = \left(\begin{array}{cc}\frac{h_4\sqrt{P}}{|h_2|}& \frac{h_3\sqrt{P}}{|h_3|}\\
\frac{h_5'\sqrt{\alpha P}}{|h_1'|}\frac{h_2\sqrt{P}}{|h_2|}& \frac{h_5'\sqrt{\alpha P}}{|h_1'|}\frac{h_1\sqrt{P}}{|h_3|}\end{array}\right).
\end{equation}
\begin{equation}
\xi = 1 + \frac{|h_5'|^2}{|h_1'|^2}\alpha P.
\end{equation}
\begin{itemize}
\item Option 1:
\begin{equation}
_2\mathrm{SNR}^1_1=\frac{\xi\gamma_{11} + \gamma_{12} + \gamma_b}{\xi \gamma_{21} +  \gamma_{22} + \xi}.
\end{equation}
\begin{equation}
_2\mathrm{SNR}^1_2=\gamma_{21} + \frac{\gamma_{22}}{\xi}.
\end{equation}
\item Option 2:
\begin{equation}
_2\mathrm{SNR}^2_2=\frac{\xi\gamma_{21} + \gamma_{22} + \gamma_b}{\xi \gamma_{11} +  \gamma_{12} + \xi}
\end{equation}
\end{itemize}
\begin{equation}
\begin{array}{lll}
\left\{\begin{array}{l}R_1 \leq \C\left(\min\left(_S\mathrm{SNR}_1, _2\mathrm{SNR}_1^1 \right)\right)\\R_2 \leq \C\left(_2\mathrm{SNR}_2^1\right)\end{array}\right.&
\mbox{or}&
\left\{\begin{array}{l}R_1 \leq \C\left(_S\mathrm{SNR}_1\right)\\R_2 \leq \C\left(_2\mathrm{SNR}_2^2\right)\end{array}\right.
\end{array}
\end{equation}
\subsection{$S_3$}
\begin{equation}
y_2 = \frac{h_3\sqrt{P}}{|h_1|}x_1 + z_2.
\end{equation}
\begin{equation}
y_R = \frac{h_1\sqrt{P}}{|h_1|}x_1 + z_R.
\end{equation}
\begin{equation}
\alpha = \frac 1{P + \sigma^2}.
\end{equation}
\begin{equation}\begin{array}{lll}
y_1' &=& \frac{h_2'\sqrt{\alpha P}}{|h_2'|}y_R + z_1'\\
&=&
\frac{h_2'\sqrt{\alpha P}}{|h_2'|}\frac{h_1\sqrt{P}}{|h_1|}x_1 + \frac{h_2'\sqrt{\alpha P}}{|h_2'|}z_R + z_1'.
\end{array}\end{equation}
\begin{equation}
_1\mathrm{SNR}_1 = \frac{\alpha P^2}{\alpha P\sigma^2 + \sigma^2}.
\end{equation}

\begin{equation}\begin{array}{lll}
y_2' &=& \frac{h_5'\sqrt{\alpha P}}{|h_2'|}y_R + \frac{h_3'\sqrt{P}}{|h_3'|}x_2 + z_2'\\
&=&
\frac{h_5'\sqrt{\alpha P}}{|h_2'|}\frac{h_1\sqrt{P}}{|h_1|}x_1 + \frac{h_5'\sqrt{\alpha P}}{|h_2'|}z_R + \frac{h_3'\sqrt{P}}{|h_3'|}x_2 + z_2',
\end{array}\end{equation}

\begin{equation}
\mathbf{H} = \left(\begin{array}{cc}\frac{h_3\sqrt{P}}{|h_1|}& 0\\
\frac{h_5'\sqrt{\alpha P}}{|h_2'|}\frac{h_1\sqrt{P}}{|h_1|}& \frac{h_3'\sqrt{P}}{|h_3'|}\end{array}\right).
\end{equation}
\begin{equation}
\xi = 1 + \frac{|h_5'|^2}{|h_2'|^2}\alpha P.
\end{equation}

\begin{itemize}
\item Option 1:
\begin{equation}
_2\mathrm{SNR}^1_2=\frac{\xi\gamma_{21} + \gamma_{22} + \gamma_b}{\xi \gamma_{11} +  \gamma_{12} + \xi}
\end{equation}
\item Option 2:
\begin{equation}
_2\mathrm{SNR}^2_1=\frac{\xi\gamma_{11} + \gamma_{12} + \gamma_b}{\xi \gamma_{21} +  \gamma_{22} + \xi}.
\end{equation}
\begin{equation}
_2\mathrm{SNR}^2_2=\frac{\gamma_{22}}{\xi}.
\end{equation}
\end{itemize}
\begin{equation}
\begin{array}{lll}
\left\{\begin{array}{l}R_1 \leq \C\left(_1\mathrm{SNR}_1\right)\\R_2 \leq \C\left(_2\mathrm{SNR}_2^1\right)\end{array}\right.&
\mbox{or}&
\left\{\begin{array}{l}R_1 \leq \C\left(\min\left(_1\mathrm{SNR}_1, _2\mathrm{SNR}_1^2 \right)\right)\\R_2 \leq \C\left(_2\mathrm{SNR}_2^2\right)\end{array}\right.
\end{array}
\end{equation}

\subsection{$S_4$} 
\begin{equation}
y_F = \frac{h_3\sqrt{ P}}{|h_3|}x_2 + z_F.
\end{equation}
\begin{equation}
y_R = \frac{h_2\sqrt{P}}{|h_2|}x_1 + \frac{h_5\sqrt{P}}{|h_3|}x_2 + z_R.
\end{equation}
\begin{equation}
\alpha = \frac 1{P + \frac{|h_5|^2}{|h_3|^2}P + \sigma^2}.
\end{equation}
\begin{equation}\begin{array}{lll}
y_F' &=& \frac{h_1'\sqrt{\alpha P}}{|h_1'|}y_R + \frac{h_3\sqrt{\beta P}}{|h_3|}x_2 + z_F'\\
&=&
\frac{h_1'\sqrt{\alpha P}}{|h_1'|}\frac{h_2\sqrt{P}}{|h_2|}x_1 + \frac{h_1'\sqrt{\alpha P}}{|h_1'|}\frac{h_5\sqrt{P}}{|h_3|}x_2 + \frac{h_1'\sqrt{\alpha P}}{|h_1'|}z_R + \frac{h_3\sqrt{\beta P}}{|h_3|}x_2 + z_F'.
\end{array}\end{equation}
\begin{equation}
\mathbf{H} = \left(\begin{array}{cc}0& \frac{h_3\sqrt{P}}{|h_3|}\\
\frac{h_1'\sqrt{\alpha P}}{|h_1'|}\frac{h_2\sqrt{P}}{|h_2|}& \frac{h_1'\sqrt{\alpha P}}{|h_1'|}\frac{h_5\sqrt{P}}{|h_3|} + \frac{h_3\sqrt{\beta P}}{|h_3|}\end{array}\right).
\end{equation}
\begin{equation}
\xi = 1 + \alpha P.
\end{equation}
\begin{itemize}
\item Option 1:
\begin{equation}
_S\mathrm{SNR}^1_1=\frac{\xi\gamma_{11} + \gamma_{12} + \gamma_b}{\xi \gamma_{21} +  \gamma_{22} + \xi}.
\end{equation}
\begin{equation}
_S\mathrm{SNR}^1_2=\gamma_{21} + \frac{\gamma_{22}}{\xi}.
\end{equation}
\item Option 2:
\begin{equation}
_S\mathrm{SNR}^2_2=\frac{\xi\gamma_{21} + \gamma_{22} + \gamma_b}{\xi \gamma_{11} +  \gamma_{12} + \xi}
\end{equation}
\begin{equation}
_S\mathrm{SNR}^2_1=\gamma_{11} + \frac{\gamma_{12}}{\xi}.
\end{equation}
\end{itemize}
\begin{equation}
\begin{array}{lll}
\left\{\begin{array}{l}R_1 \leq \C\left(_S\mathrm{SNR}_1^1\right)\\R_2 \leq \C\left(_S\mathrm{SNR}_2^1\right)\end{array}\right.&
\mbox{or}&
\left\{\begin{array}{l}R_1 \leq \C\left(_S\mathrm{SNR}_1^2\right)\\R_2 \leq \C\left(_S\mathrm{SNR}_2^2\right)\end{array}\right.
\end{array}
\end{equation}

\bibliographystyle{ieeetr}
\bibliography{ref}

\begin{thebibliography}{10}

\bibitem{vanet_application}
C.~E. Palazzi, M.~Roccetti, and S.~Ferretti, ``An intervehicular communication
  architecture for safety and entertainment,'' {\em IEEE Trans. on Intelligent
  Transportation Systems}, vol.~11, no.~1, pp.~90--99, 2010.

\bibitem{v2v_v2i}
Q.~Wang, P.~Fan, and K.~Letaief, ``On the joint $\mathrm{V2I}$ and
  $\mathrm{V2V}$ scheduling for cooperative vanets with network coding,'' {\em
  IEEE Trans. on Veh. Tech.}, vol.~61, no.~1, pp.~62--73, 2012.

\bibitem{routing_survey}
S.~M. Bilal, C.~J. Bernardos, and C.~Guerrero, ``Position-based routing in
  vehicular networks: A survey,'' {\em Journal of Network and Computer
  Applications}, vol.~36, no.~2, pp.~685 -- 697, 2013.

\bibitem{sadv}
Y.~Ding and L.~Xiao, ``Sadv: Static-node-assisted adaptive data dissemination
  in vehicular networks,'' {\em IEEE Trans. on Veh. Tech.}, vol.~59, no.~5,
  pp.~2445--2455, 2010.

\bibitem{multiflow_chanthai}
C.~D.~T. Thai, P.~Popovski, M.~Kaneko, and E.~de~Carvalho, ``Multi-flow
  scheduling for coordinated direct and relayed users in cellular systems,''
  {\em IEEE Trans. on Commun.}, vol.~61, no.~2, pp.~669--678, Feb. 2013.

\bibitem{overhearing_fansun_comlett}
F.~Sun, T.~Kim, A.~Paulraj, E.~de~Carvalho, and P.~Popovski, ``Cell-edge
  multi-user relaying with overhearing,'' {\em IEEE Commun. Lett.}, vol.~17,
  no.~6, pp.~1--4, Jun. 2013.

\bibitem{dof_fansun_spawc}
F.~Sun and E.~De~Carvalho, ``Degrees of freedom of asymmetrical multi-way relay
  networks,'' in {\em Signal Process. Advances in Wireless Commun. (SPAWC),
  2011 IEEE 12th International Workshop on}, pp.~531--535, June 2011.

\bibitem{fourway_huaping_spawc}
H.~Liu, F.~Sun, E.~de~Carvalho, P.~Popovski, H.~Thomsen, and Y.~Zhao, ``Mimo
  four-way relaying,'' in {\em Signal Process. Advances in Wireless Commun.
  (SPAWC), 2013 IEEE 14th Workshop on}, pp.~46--50, June 2013.

\bibitem{fourway_huaping_arxiv}
H.~Liu, P.~Popovski, E.~de~Carvalho, Y.~Zhao, F.~Sun, and C.~D.~T. Thai,
  ``Transmission schemes for four-way relaying in wireless cellular systems,''
  {\em http://arxiv.org/pdf/1209.5829v1.pdf}.

\bibitem{fourway_huaping_lett}
H.~Liu, P.~Popovski, E.~de~Carvalho, Y.~Zhao, and F.~Sun, ``Four-way relaying
  in wireless cellular systems,'' {\em IEEE Wireless Commun. Lett.}, vol.~2,
  pp.~403--406, August 2013.

\bibitem{dmt_chanthai}
C.~D.~T. Thai, P.~Popovski, E.~de~Carvalho, and F.~Sun,
  ``Diversity-multiplexing trade-off for coordinated direct and relay
  schemes,'' {\em IEEE Trans. Wireless Commun.}, vol.~12, no.~7,
  pp.~3289--3299, Jul. 2013.

\bibitem{backlog3}
Y.~Ma, ``Rate-maximization scheduling for downlink ofdma with long term rate
  proportional fairness,'' in {\em Commun., 2008. ICC '08. IEEE International
  Conference on}, pp.~3480--3484, 2008.

\bibitem{backlog2}
W.~Cheng, M.~Ghogho, Q.~Huang, D.~Ma, and J.~Wei, ``Maximizing the sum-rate of
  amplify-and-forward two-way relaying networks,'' {\em IEEE Signal Process.
  Lett.}, vol.~18, no.~11, pp.~635--638, 2011.

\bibitem{pf1}
H.~Cho and J.~Andrews, ``Resource-redistributive opportunistic scheduling for
  wireless systems,'' {\em IEEE Trans. on Wireless Commun.}, vol.~8, no.~7,
  pp.~3510--3522, 2009.

\bibitem{backlog}
M.~Andrews and L.~Zhang, ``Scheduling algorithms for multicarrier wireless data
  systems,'' {\em IEEE/ACM Trans. on Networking}, vol.~19, no.~2, pp.~447--455,
  2011.

\bibitem{femto_uplink}
V.~Chandrasekhar and J.~Andrews, ``Uplink capacity and interference avoidance
  for two-tier femtocell networks,'' {\em IEEE Trans. on Wireless Commun.},
  vol.~8, no.~7, pp.~3498--3509, 2009.

\bibitem{mmse_fansun_siglett1}
F.~Sun and E.~de~Carvalho, ``A leakage-based mmse beamforming design for a mimo
  interference channel,'' {\em IEEE Signal Process. Lett.}, vol.~19, no.~6,
  2012.

\bibitem{mmse_fansun_siglett2}
T.~M. Kim, F.~Sun, and A.~J. Paulraj, ``Low-complexity mmse precoding for
  coordinated multipoint with per-antenna power constraint,'' {\em IEEE Signal
  Process. Lett.}, vol.~20, no.~4, 2013.

\end{thebibliography}

\end{document}